\documentclass[prb, twocolumn]{revtex4-1}
\usepackage{amsmath,amssymb,bm}
\usepackage{graphicx}
\usepackage{epstopdf}
\usepackage{latexsym}
\usepackage{subfigure}
\usepackage{color}
\usepackage{natbib}
\usepackage{hyperref}
\usepackage{braket}
\hypersetup{
  colorlinks,
  citecolor=magenta,
  linkcolor=blue,
  urlcolor=blue}
      
\newcommand{\bsun}{${\cal N}$-$\bar {\cal N}$}

\begin{document}

\title{Marshall-positive SU($N$) quantum spin systems and classical loop
  models: A practical strategy to design ``sign-problem''-free spin Hamiltonians}
\author{Ribhu K. Kaul}
\affiliation{Department of Physics \& Astronomy, University of Kentucky, Lexington, KY-40506-0055}
\begin{abstract}
We consider bipartite SU($N$) spin Hamiltonians with a fundamental
representation on one sublattice and a conjugate to fundamental on
the other sublattice.  By mapping these antiferromagnets to certain classical
loop models in one higher dimension, we provide a practical strategy to
write down a large family of SU($N$)
symmetric spin Hamiltonians that satisfy Marshall's sign condition. 
This family includes all previously
known sign-free SU($N$) spin models in this representation and in addition provides a large
set of new models that are Marshall positive and
can hence be studied efficiently with quantum Monte Carlo methods. As an
application of our idea to the square lattice, we show that in addition to Sandvik's
$Q$-term, there is an independent non-trivial four-spin $R$-term that is
sign-free. Using numerical simulations, we show how the $R$-term
provides a new route to the study of quantum criticality of N\'eel order.
\end{abstract}
\maketitle

\section{Introduction}

The study of the ground states of lattice spin Hamiltonians has a long
history in physics.~\cite{bethe1931:chain} Due to its relevance to
quantum magnetism in solid state materials, the study is currently one
of the cornerstones of modern condensed matter physics.~\cite{balents2010:spliq}
 The scarcity of controlled analytic
 solutions of spin models has spurred the development of a wide array of
 sophisticated numerical approaches.~\cite{scholl2005:dmrg,gelfand2000:ser_exp,laflorencie2004:ed}  Quantum Monte Carlo (QMC) is often the method of choice
 for unbiased studies of large higher dimensional ($d>1$) quantum spin
 systems.~\cite{sandvik2010:vietri} In practice this capability is
 restricted to models that do not suffer from the notorious
 sign-problem.~\cite{henelius2000:sign,nyfeler2008:frust} The absence of this problem is guaranteed in standard
 world line methods only if
 the Marshall sign rule~\cite{auerbach1994:intelc} ($\langle \alpha| H |\beta \rangle <0$ for $\alpha \neq \beta$) is satisfied in a convenient local basis. 
Surprisingly, at the current time, there is no systematic knowledge of the extent of
 Marshall-positive spin Hamiltonians.
Such an understanding would
 clearly be of great practical and conceptual value, since 
 Marshall-positive Hamiltonians constitute a large fraction of the valuable examples of higher-dimensional models that can be simulated
 on a classical computer in polynomial time. \cite{terhal2008:stoquastic,kaul2013:qmc}

\begin{figure}[!t]
\centerline{\includegraphics[angle=0,width=1.0\columnwidth]{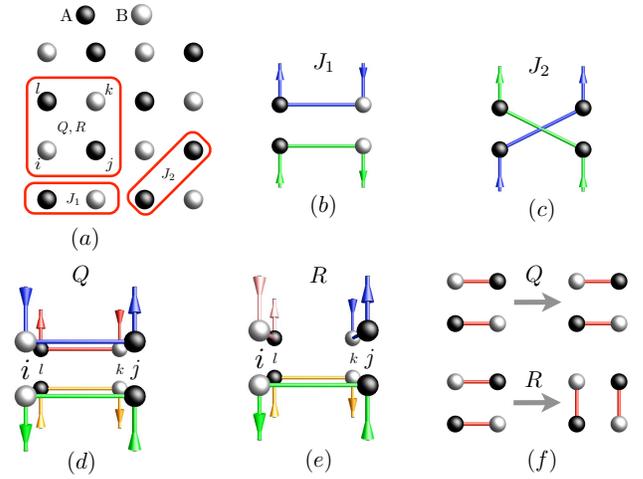}}
\caption{Cartoon illustration of the various Marshall positive interaction
  terms studied in the paper. (a) shows the decomposition of the
  bipartite square lattice into A and B sublattices, and the groups
  of sites that the various interactions act on. (b-e) cartoons of the
  different interactions acting on the basis states, the initial state is
  shown at the bottom and the final state is shown on top. (b) shows the
  $J_1$ interaction that acts between sites on different
  sublattices. (c) shows the $J_2$ interaction that acts on sites on
  the same sublattice. (d) shows an example of a $Q$ interaction and (e) shows
 an example of an $R$ interaction that both act on elementary plaquettes of the
  square lattice. Colors of the loop represent the $N$ colors of
  SU($N$). Note that loops meeting at a vertex may have the same color
  and that the loops always travel in opposite directions on the A and B sublattices. (f) Cartoon illustration of singlet rearrangements effected
  by $R$ and $Q$ terms, see discussion on Eq.~(\ref{eq:RQsing}). }
\label{fig:intro}
\end{figure}

Here we will address this issue for a
specific but important subset of models, SU($N$) quantum spin
Hamiltonians on the square
lattice, which have spins that transform as the fundamental
representation (${\cal N}$) on the A sublattice and the conjugate to
fundamental ($\bar {\cal N}$) on
the B sublattice, Fig.~\ref{fig:intro}(a).~\cite{affleck1985:lgN}
 Let us call the set of all such Hamiltonians
\bsun{}. Note that for $N=2$ the fundamental representation is
self-conjugate so this class includes all SU(2) symmetric Hamiltonians.  
\bsun{} includes both
models that are widely believed to have no solution to the sign
problem (e.g. the $J_1$-$J_2$ Heisenberg model with $J_1,J_2$ anti-ferro~\cite{nakamura1993:j1j2qmc}) and others for which the absence of the sign problem is well
known. The simplest interaction in \bsun{} was introduced by Affleck, \cite{affleck1985:lgN}
\begin{equation} 
\label{eq:affleck}
H^{(N)}_{J_1} =-  J_1 T^a_i \cdot T^{a*}_j,
\end{equation} 
where $i$ and $j$ are on different sublattices of the
bipartition and $T^a$ are the SU($N$) generators in the fundamental
representation (sum on $a$ is implied). It happens that this interaction is
Marshall positive and can hence be simulated efficiently with QMC.~\cite{harada2003:sun} We note
that the familiar SU(2) Heisenberg antiferromagnet, $H^{(2)}_{J_1} = J_1 {\vec S}_i \cdot
{\vec S}_j$, a special case of Eq.~(\ref{eq:affleck}) with $N=2$,  was
the original model for which Marshall proved his theorem.~\cite{marshall1955:mst} This fact has been exploited to study it using QMC on various ordered and
disordered bipartite lattices for the last three decades (see for
example the
bibliography in~Ref. [\onlinecite{sandvik2010:vietri}]).
In an important breakthrough, Sandvik found that in addition to the
two spin interactions, a four-spin plaquette $Q$-term,~\cite{sandvik2007:deconf}
\begin{equation}
\label{eq:Q}
H^{(2)}_Q
= -Q \left ( {\vec S}_i \cdot  {\vec S}_j - \frac{1}{4}
\right ) \left ( {\vec S}_k \cdot  {\vec S}_l - \frac{1}{4}
\right ) + i \leftrightarrow k,
\end{equation} 
also satisfied Marshall sign
condition (see Fig.~\ref{fig:intro}(a) for site labeling). The sign-free $J_1$-$Q$ hamiltonian
has since been studied extensively using QMC~\cite{melko2008:jq,jiang2008:first,harada2013:deconf,sandvik2010:logs,banerjee2010:log}.
The discovery of the $Q$-idea has lead to the proposal and study of a
number of extensions, 
including generalizations for $N>2$ along the lines of Eq.~(\ref{eq:affleck}).~\cite{sen2010:first,pujari2013:hc,lou2009:sun,kaul2011:su34,banerjee2010:su3,block2013:fate}
The discovery of the $Q$-term and its popularity in
numerical studies begs the questions:
Are there other \bsun{} models that satisfy Marshall's sign rule? What
is the full extent of these sign-free models?

\section{General sign free \bsun{} models} 

\label{sec:design}

\begin{figure}[!t]
\centerline{\includegraphics[width=3.6in,trim=180 80 200 80,clip=true]{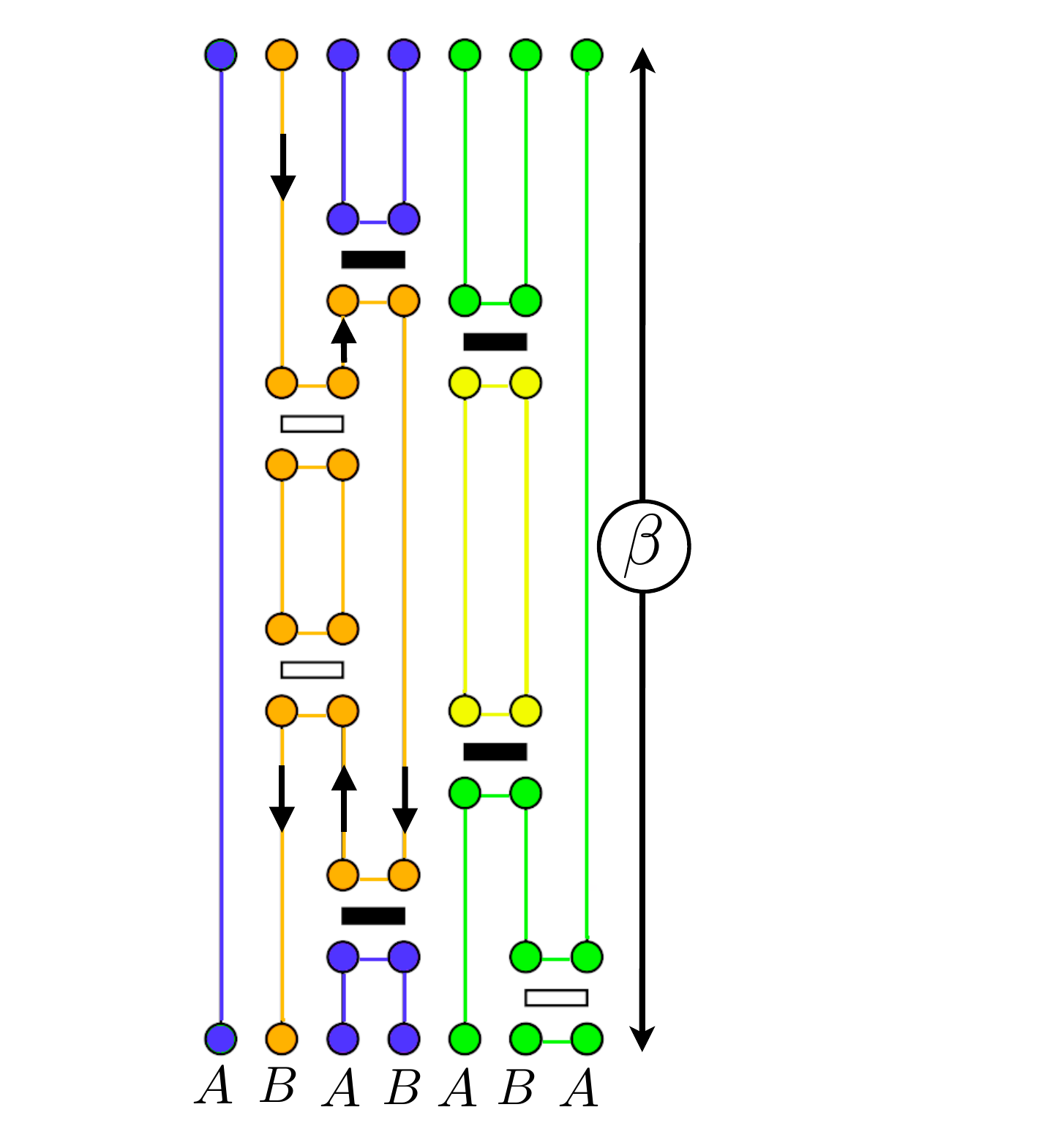}}
\caption{An illustration of a path-integral or SSE contribution to the
partition function of an SU($N$) anti-ferromagnet with only $J_1$
interactions with 7 sites at a temperature $\beta$, which may be
viewed as a higher dimensional closed packed loop model. The orientation of
one particular loop is followed with
black arrows to show how it always travels in imaginary time in opposite
directions on opposite sublattices (up on A and down on B). This
orientation is preserved by the $J_1$ interaction as well as all other
SU($N$) invariant interactions. An SU($N$) antiferromagnet will have $N$
different color assignments to the loops. }
\label{fig:jon}
\end{figure}

We now show that it is easiest to address these questions by considering
the structure of the imaginary time statistical mechanics ($Z={\rm
  Tr}[e^{-\beta H}]$) generated by the \bsun{}
Hamiltonians. Each site on the bipartite lattice has $N$ states. Our goal is to write
down sign-positive model interactions that are invariant
under rotation by $U$ (an SU($N$) matrix) on all the A sublattice
spins and rotation $U^*$ on all the B sublattice spins. Let us begin
by reviewing how this works for a two-spin interaction, which we can write in the form $\Gamma_{\alpha\beta\gamma\eta}|\alpha\beta\rangle
\langle \gamma \eta |$.  To preserve SU($N$) invariance, the spin
indices on the
sites have to be paired up. If the two sites are on opposite
sublattices the only interaction is
$H^{(N)}_{J_1} = -\frac{J_1}{N}\sum_{\alpha \beta}  |\alpha \alpha \rangle \langle \beta \beta|$
which can be shown to be unitary equivalent to Eq.~(\ref{eq:affleck})
up to a constant. Between sites on the same sublattice we must have
$H^{(N)}_{J_2} = -\frac{J_2}{N}\sum_{\alpha \beta}  |\alpha \beta \rangle
\langle \beta \alpha|$ ~\cite{kaul2012:j1j2, kaul2012:bilayer} (for $N=2$ this is a ferromagnetic Heisenberg
interaction between sites on the same sublattice). The effect of
these two-spin terms on the imaginary time evolution of the basis states can be
represented by the diagrams shown in Fig.~\ref{fig:intro}(b), (c). It becomes clear that
the quantum statistical mechanics of models in \bsun{} is equivalent
to the classical statistical mechanics of specific tightly packed
loop models of $N$ colors in one higher dimension. The staggered
representations of SU($N$) appear in the loop picture by requiring 
the orientation of a given loop to be such that it travels (in
imaginary time) in
opposite directions on opposite sublattices. Stated explicitly, if the
loop travels up when
its on an A site, it must always do so on all other A sites and it must always travel down when its on a B site; this
property is illustrated for a 1-d system with $J_1$ interactions in
Fig.~\ref{fig:jon}. SU($N$) invariant Hamiltonian operators must reconnect the loops
without termination and preserve the directionality of the loops.   Avoiding the sign problem only
requires that the weight associated with a reconnection of loops be
positive (this is Marshall's sign condition).

Let us  apply these ideas to design Marshall-positive four-spin Hamiltonian terms
acting on the sites of an elementary plaquette. The simplest such connection is
shown in Fig.~\ref{fig:intro}(d), writing this as a Hamiltonian
operator, it is $H^{(N)}_Q =-\frac{Q}{N^2}\sum_{\alpha\beta\gamma\eta} \left ( |\alpha
\alpha \beta\beta \rangle \langle \gamma\gamma \eta \eta |+ |\beta
\alpha \alpha\beta \rangle \langle \eta\gamma \gamma \eta | \right )$  (the bra
and kets
are labeled by the spin indices at sites, $ijkl$). It is easy to prove
that for $N=2$ this
interaction is unitarily equivalent to the $Q$ term in
Eq.~(\ref{eq:Q}).  Interestingly, an independent four-spin interaction can be achieved by
re-connecting the loops differently as shown in
Fig.~\ref{fig:intro}(e). In the bra-ket notation the new Hamiltonian
is,
\begin{equation}
\label{eq:RN}
H^{(N)}_R =-\frac{R}{N^2}\sum_{\alpha\beta\gamma\eta} \left ( |\alpha
\alpha \beta\beta \rangle \langle \eta\gamma \gamma \eta | + |\beta
\alpha \alpha\beta \rangle \langle \gamma\gamma \eta \eta |\right ),
\end{equation}
Written this way, the Marshall positivity of the
$R$-term is obvious.  We point out however
that, unlike the $Q$ term which is a product of two Marshall-positive
$J_1$ terms, in the usual language
the $R$ interaction's positivity is non-trivial. Indeed, for $N=2$, our new term may be written with
familiar $S=1/2$ operators on each elementary plaquette (labeling the sites $ijkl$ cyclically as shown in Fig.~\ref{fig:intro}(a))
\begin{widetext}
\begin{equation}
\label{eq:Rint_spin}
H^{(2)}_R= R\left(\left( \vec S_i\cdot \vec S_k -\frac{1}{4}\right) \left( \vec S_l\cdot \vec S_j -\frac{1}{4}\right)- \left( \vec S_i\cdot \vec S_l -\frac{1}{4}\right) \left( \vec S_k\cdot \vec S_j -\frac{1}{4}\right) - \left( \vec S_i\cdot \vec S_j -\frac{1}{4}\right) \left( \vec S_k\cdot \vec S_l -\frac{1}{4}\right)\right).
\end{equation}
\end{widetext}
In contrast to the $Q$-term, the first term contains dot products between spins on sites {\em on
  the same} sublattice. Naively the  first term would seem to make
the $R$-interaction violate the Marshall rule, and indeed the first term {\em by itself} would. Remarkably however when all three terms of the $R$-interaction are taken together the offending matrix elements exactly cancel (after one does the usual Marshall rotation by an angle of $\pi$ about the $z$-axis for the spins on one of the sublattices). 
We can
re-write the $Q$ and $R$ terms in a way in which their physics is more transparent,
\begin{eqnarray}
\label{eq:RQsing}
H^{(N)}_{Q}&=&- Q \left (|S_{ij}S_{kl}\rangle \langle S_{ij}S_{kl}| +  |S_{il}S_{kj}\rangle \langle S_{il}S_{kj}| \right )\nonumber \\
H^{(N)}_{R}&=&- R \left (|S_{ij}S_{kl}\rangle \langle S_{il}S_{kj}| +
  |S_{il}S_{kj}\rangle  \langle S_{ij}S_{kl}| \right )\nonumber \\
\end{eqnarray} 
where $|S_{ij}\rangle = \frac{1}{\sqrt{N}}\sum_\alpha
|\alpha_i\alpha_j\rangle$ is the SU($N$) singlet (refer to
Fig.~\ref{fig:intro}(a) for $ijkl$ labeling). Now the physical
distinction between these two terms is apparent, see Fig.~\ref{fig:intro}(f): The $Q$-interaction
is a diagonal attractive term between two neighboring parallel
singlet, whereas the $R$-interaction is an off-diagonal term that
causes neighboring  $x$-oriented singlets to become $y$-oriented
singlets on an elementary plaquette.
We note here that the $Q$ and $R$ are very reminiscent
of the diagonal and off-diagonal terms respectively of the original
quantum dimer model,~\cite{rokhsar1988:qdm} but defined for spin models. Indeed in the
$N\rightarrow \infty$ mapping of the SU($N$) antiferromagnet to the quantum dimer model,~\cite{read1989:nucphysB}
these are precisely the terms they would become.

\begin{figure}[!t]
\centerline{\includegraphics[width=2.4in,trim=0 0 0 0,clip=true]{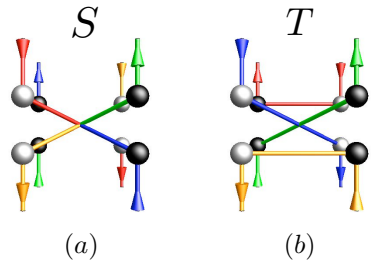}}
\caption{For completeness we show here the diagrams that correspond to
  two more sign-free four-spin
  interactions, in addition to those shown in
  Fig.~\ref{fig:intro}(d,e).  Taken together, these four terms are the complete
set of sign-free four-spin interactions that can be obtained from our construction.}
\label{fig:st}
\end{figure}

We note here that both $Q$ and $R$ act within the total spin zero sector of the
four sites. For completeness, two additional sign-free 4-spin
interactions ($S$ and $T$) that act on
higher-spin sectors are shown in Fig.~\ref{fig:st}. Analogous to our discussion above for
2- and 4-spin interactions, we can systematically enumerate the
6-, 8-, or even higher spin
interactions and can be carried out on any bipartite lattice by constructing loop reconfigurations of the kind shown
in Fig.~\ref{fig:intro} (b-e) with the desired sets of sites. The
study of this
large family of Marshall positive Hamiltonians is an exciting direction for future work. In the
remainder of this manuscript we
present a study of the new $R$ interaction.

\begin{figure}[t!]
\centerline{\includegraphics[angle=0,width=1.0\columnwidth]{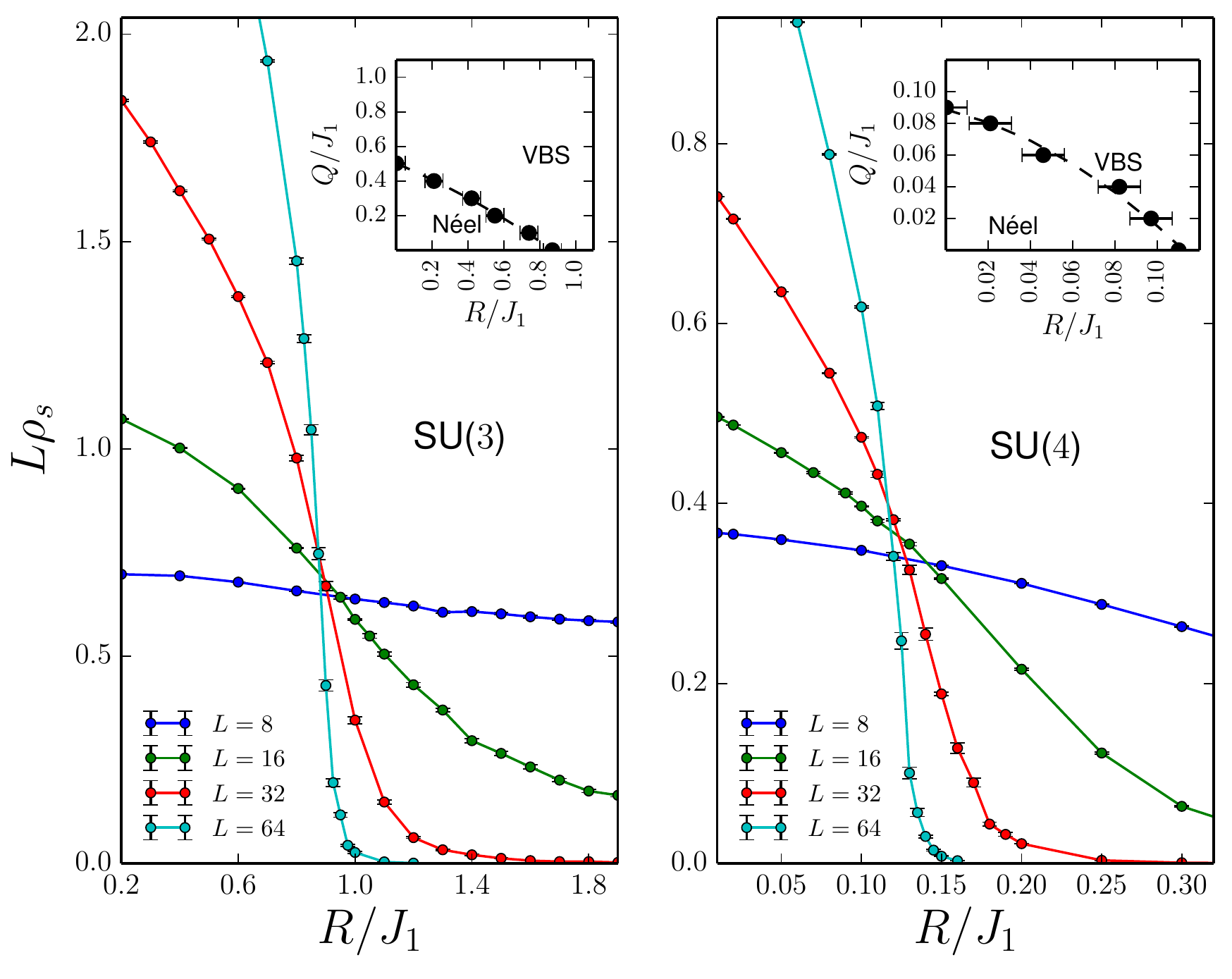}}
\caption{Magnetic phase transitions from the crossing of $\rho_s$
  (stiffness) data shown for SU(3) and SU(4). In the main
  panels data for $L\rho_s$ is shown for the $J_1$-$R$ model  as a function of $R/J_1$. Clear evidence for a
  crossing is found which implies the existence of a critical point
  where magnetic order is destroyed.  The insets show
  the magnetic phase boundaries for the $J_1$-$Q$-$R$ model inferred from the kind
  of data shown in the main panel with $Q\neq 0$ (not shown). We have
  chosen $\beta=L$ and $J_1=1$ here.}
\label{fig:rho_pd}
\end{figure}

\begin{figure}[!t]
\centerline{\includegraphics[angle=0,width=1.0\columnwidth]{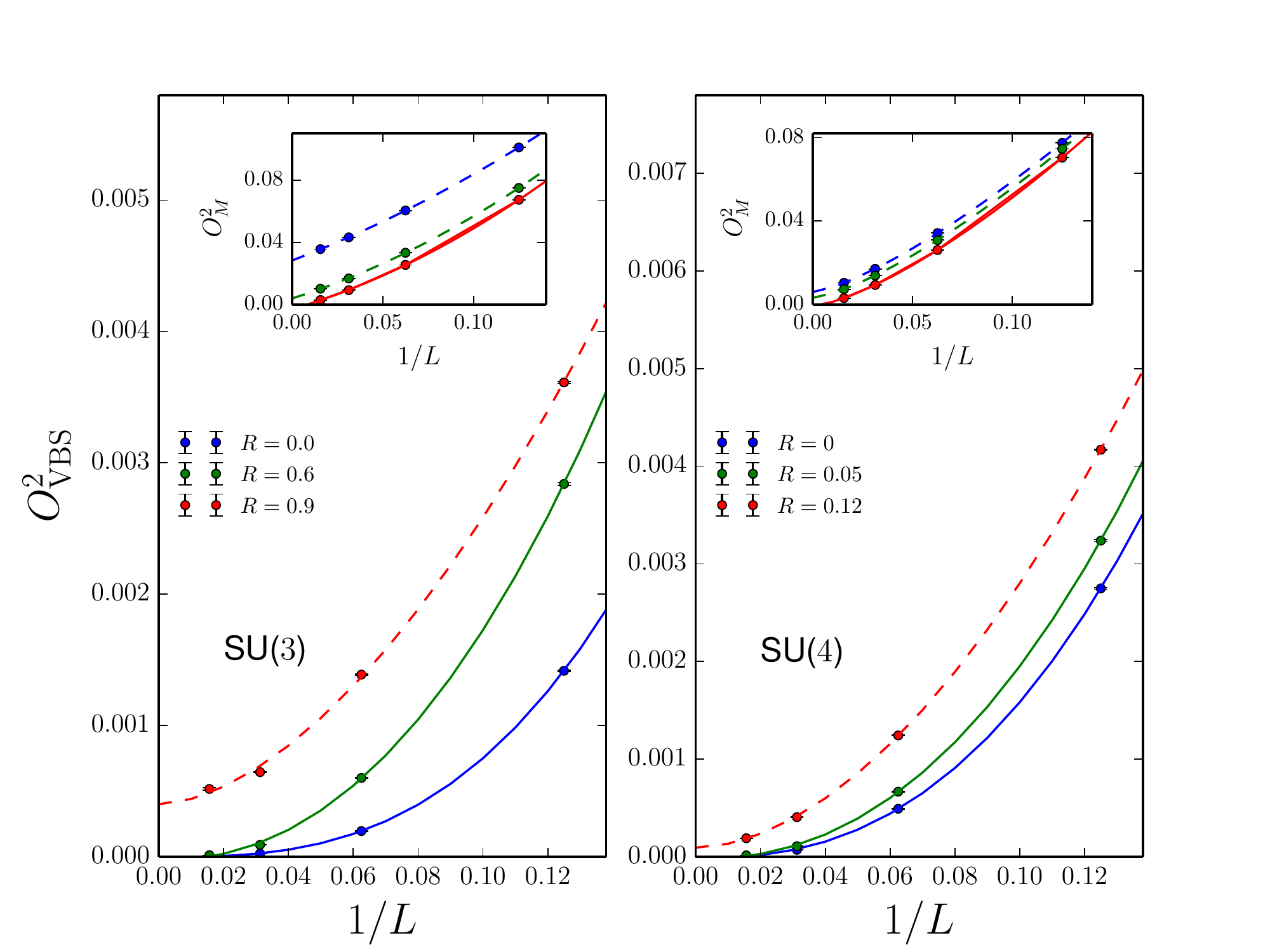}}
\caption{Representative VBS (main panels) and magnetic (insets) order parameters for SU(3) and
  SU(4) for $R=0$ and on either side close to the critical points (the
  order parameters are defined in the SM). All fits through the QMC data points are shown only as a
  guide to the eye. The dashed (solid) lines correspond to cases where
  finite (zero)
  order parameters are obtained in the thermodynamic limit. }
\label{fig:fss}
\end{figure}

\section{Numerical Simulations} 

Since the SU($N$) symmetric $J_1$-$R$
model on the square lattice does not suffer
from a sign problem we are able to study it on large system sizes and
at low
temperatures.
 Here we use the stochastic series
expansion (SSE) method for which we have efficient loop updates for the
quantum Monte Carlo.~\cite{sandvik2010:vietri}

It is well known that the $J_1$ only model has N\'eel order for
$N=2,3,4$.~\cite{harada2003:sun} We first ask whether the new $R$ interaction can destroy this
N\'eel order in the $J_1$-$R$ model. While this is not the case for
$N=2$ (see Supplemental Materials), clear indications for crossings of $\beta \rho_s = \langle
W^2 \rangle$ (the average square of the winding number of the loops), which signal
the destruction of magnetic order, are shown for SU(3) and SU(4) in
Fig.~\ref{fig:rho_pd}. It has been previously shown that the $Q$
interaction can also destroy the N\'eel order for $N=3,4$, giving rise to a
VBS.~\cite{lou2009:sun} For completeness, the insets  in Fig.~\ref{fig:rho_pd} show the phase diagram of
the $J_1$-$Q$-$R$ model to connect our work with previous work on the
$J_1$-$Q$ model.

The natural candidate for the large $R$ non-magnetic state is a valence-bond
solid. Finite-size scaling of the VBS and N\'eel order parameters
($O_{\rm VBS}$ and $O_{M}$)
close to the critical point
confirm this expectation as shown in Fig.~\ref{fig:fss}. The
definition of the order parameters are standard, they are included in the SM
for completeness. It is found
that the N\'eel order parameters turn off at the same time the VBS
order parameters first come on,
consistent with the observation of such behavior in the $J$-$Q$
model~\cite{lou2009:sun} and with the deconfined criticality
scenario.~\cite{senthil2004:science} An in depth study of the critical
behavior is beyond the scope of the current work and will be presented
elsewhere.

\begin{figure}[t!]
\centerline{\includegraphics[angle=0,width=1.0\columnwidth]{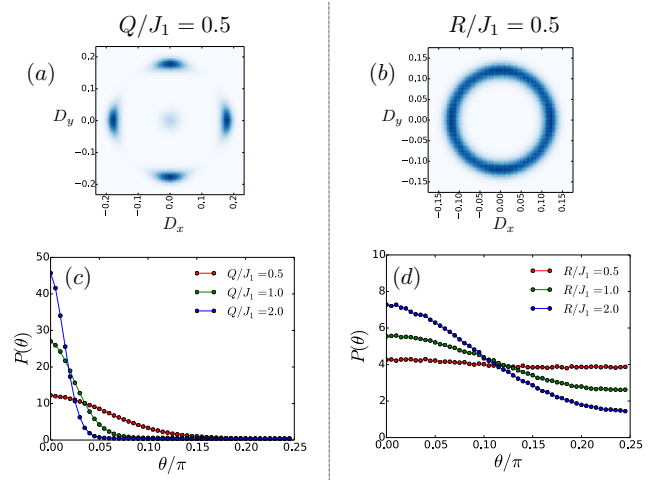}}
\caption{Comparison and contrast of the VBS ordering in the $J_1$-$Q$ (left) and
  $J_1$-$R$ (right) model for SU(4). (a,b) show 2-$d$ histograms of the VBS order
  parameter, $D_y$ and $D_x$, measured at equal time on a
  $64\times 64$  systems at a
coupling of $Q/J_1=0.5$ and $R/J_1=0.5$ (deep in the VBS phase for both models, see inset
of right panel of Fig.~\ref{fig:rho_pd} for phase diagrams). At the same couplings, the columnar nature of the
ordering is much more apparent in the $Q$ model as compared to the $R$
model, where a U(1) symmetry is observed. (c,d) Showing the
probability $P (\theta)$ of getting a particular angle for
the VBS order parameters on $64\times 64$ system for different
couplings. By symmetry the function repeats itself in an eight-fold way and hence we show it
only for $0<\theta<\pi/4$ (a half period). Note again the weak decay of $P$ with
$\theta$ in
the $R$-model, as compared to the $Q$-model. See text for discussion. 
}
\label{fig:hist}
\end{figure}

We note the
VBS order parameter is finite in both the columnar and
plaquette VBS states. It is now well established that the $Q$ term
favors columnar order,~\cite{lou2009:sun} consistent with its
interpretation as an attraction between neighboring parallel dimers. Since the $R$ term appears to be more like the
kinetic term in the dimer model (see Eq.~(\ref{eq:RQsing}) and discussion), it is
interesting to ask what kind of VBS ordering the $R$-term favors. In
order to make a comparison of the two, we have studied both with the VBS histogram technique.~\cite{sandvik2007:deconf,kawashima2007:sun}
A summary of our histogram study is shown in
Fig.~\ref{fig:hist}. For each configuration of our MC sample we can
calculate the value of $D_x$ (VBS order parameter with $x$-oriented
dimers Fourier transformed to $(\pi,0)$) or $D_y$ (analogously
defined) -- see SM for more details. A columnar VBS state should show a desire to be either only in
$D_x$ or only in $D_y$ at a given time, whereas a plaquette VBS would
show equal probability of being in both at any given time. We show
data for the SU(4) case for which both
$Q$ and $R$ destroy the magnetic order at a coupling ratio of order
0.1. So the data in Fig.~\ref{fig:hist} is deep in the VBS
phase for both models. Consistent with this both systems show well formed valence
bonds, as is clear by looking at the radius of the histograms in 
(a) and (b), though for an equally strong coupling the $Q$ term has a
larger amplitude. This weaker VBS order is also evident in the angular
distribution.  Here the $Q$ terms shows very strong affinity for a
columnar VBS with sharp peaks forming along the axis at $(\pm
D^0_x,0)$ and $(0,\pm D^0_y)$, as shown in (c). In contrast, the $R$ term admits a broad angular distribution even
for large value of the $R/J_1$ and big volumes (the data in (d) is
shown for $64\times 64$). As $R/J_1$ is increased even further, there
appears to be a trend towards a columnar VBS with increase of
weight around $\theta=0$, though as shown in Fig.~\ref{fig:hist} this
crossover is very slow. This is somewhat surprising since both $R$ and $Q$ destroy
N\'eel order at approximately the same value of the coupling. This new
microscopic route to quantum criticality provided by the $R$ term is an alternative to
the well-studied $J$-$Q$ model and will be useful to test the independence from
microscopic details of critical
exponents and other putative universal quantities in future studies.

\section{Summary}

To conclude, we have introduced a systematic method to generate an
infinitely large family of SU($N$)
Marshall positive Hamiltonians on bipartite lattice with multi-spin
interactions. Our understanding of this family of models is poor and
their study raises intriguing questions --
How do we choose our coupling in this large paramater space to
stabilize new phases of matter, {\em e.g.}, a spin liquid, in a
sign-positive Hamiltonian? In what sense is the
family of models we have found a complete set of Marshall positive models?
As a practical application of our idea, we provided an alternate way to see the
positivity of the popular $Q$-interaction,~\cite{sandvik2007:deconf} and discovered an independent
positive four-spin
$R$-term, which can be utilized as a new route to quantum
criticality. In a straightforward manner, our method can be extended to
classify interactions that involve more than four spins on the square
lattice as well as other
bipartite lattices in arbitrary dimensions. Our general approach
might also be useful to design Marshall positive models with symmetries different from
the \bsun{} SU($N$) model. 

\section{Acknowledgements}
 
The author thanks K. Damle, T. C. Lang,
M. A. Levin and
A. W. Sandvik for stimulating discussions, J. Demidio for his help in
preparing some figures,
NSF DMR-1056536  for partial financial support and the visitor
programs at BU for their
hospitality during the preparation of this manuscript.
Part of this work was completed while the author held
an adjunct faculty position at the TIFR.

\bibliography{/Users/rkk/OPPIE/Physics/PAPERS/BIB/career.bib}

\appendix

\section{Test Energies}

For future reference, Table~\ref{tab:qN} contains test comparisons between the energies obtained from
a SSE-QMC study and exact diagonalization on $4\times 4$ and $4 \times
6$ systems with various $J_2$-$Q$-$R$ at $N=2$, working with
$J_1=1$. The various interaction terms are described in Sec. \ref{sec:design}.

\begin{table}[h]
\begin{tabular}{||c||c|c|c|c|c|c||} 
\hline 
\hline
size & $J_2$ & $Q$ & $R$ & $\beta_{\rm QMC}$ & $E_{ex}$ & $E_{\rm QMC}$  \\
\hline
\hline
$4\times 4$ & $1$ & $0$ & $0$ &16& $-1.164574932621$ & $-1.16453(2)$\\
\hline
$4\times 4$ & $0$ & $1$ & $0$ &16& $-1.547407628767$ & $-1.54735(2)$\\
\hline
$4\times 4$ & $0$ & $0$ & $1$ &16& $-1.532604539066$ & $-1.53258(3)$\\
\hline
$4\times 4$ & $0$ & $1$ & $1$ &16& $-2.395806575537$ & $-2.39577(4)$\\
\hline
$4\times 4$ & $0$ & $1$ & $4$ &16& $-4.947607904937$ & $-4.94757(8)$\\
\hline
$6\times 4$ & $1$ & $0$ & $0$ &16& $-1.144492865177$ & $-1.14438(2)$\\
\hline
$6\times 4$ & $0$ & $1$ & $0$ &16& $-1.51911890163$ & $-1.51913(3)$\\
\hline
$6\times 4$ & $0$ & $0$ & $1$ &16& $-1.50323567768$ & $-1.50323(3)$\\
\hline
$6\times 4$ & $0$ & $1$ & $1$ &16& $-2.349607914424$ & $-2.34962(4)$\\
\hline
$6\times 4$ & $0$ & $1$ & $4$ &16& $-4.847067604114$ & $-4.84711(6)$\\
\hline
\hline 
\end{tabular}
\caption{Test comparisons of energies from exact diagonalization and
 finite-$T$ QMC studies for the SU(2) model.  Note that $J_1=1$ always. The energies reported here are per site and
on square lattices with periodic boundary conditions.}
\label{tab:qN}
\end{table}

\section{Phase diagram for $N=2$}

\begin{figure}[!h]
\centerline{\includegraphics[angle=0,width=1.0\columnwidth]{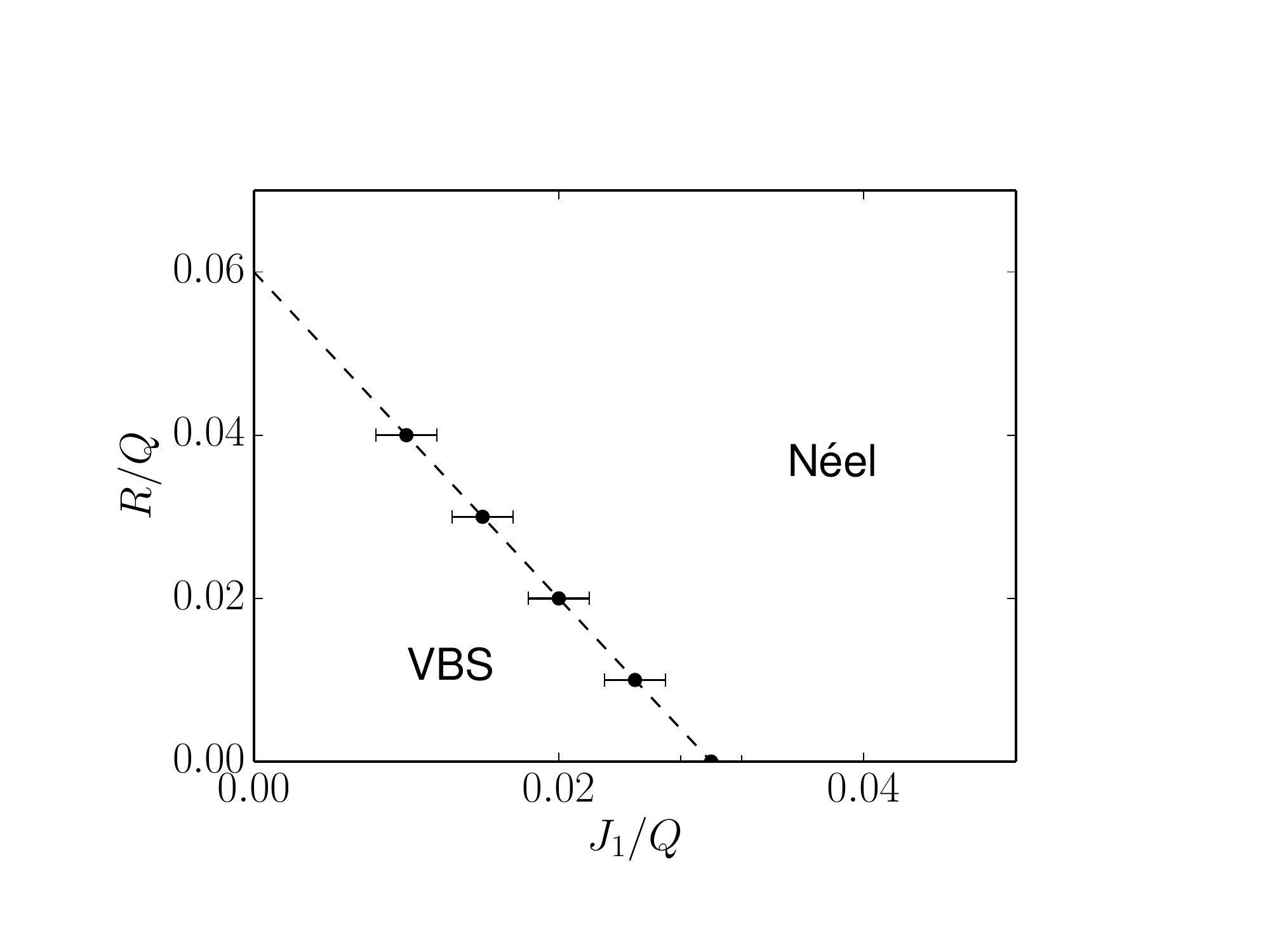}}
\caption{An approximate phase digram of the SU(2) $J_1$-$Q$-$R$ model. This graph shows the location of the crossings of the size
  $L=16$ and $L=32$ data for
 $L\rho_s$. Similar to data shown in Fig.~\ref{fig:rho_pd} but on
 smaller system sizes. }
\label{fig:su2_pd}
\end{figure}

For completeness in Fig.~\ref{fig:su2_pd} shows an approximate phase diagram
of the SU(2) $J_1$-$Q$-$R$ model, by monitoring the crossings of $L\rho_s$. It is well known that
  there are drifts in the critical point extracted
  from such crossings but that they eventually converge,~\cite{melko2008:jq,sandvik2010:logs} so this is an
  approximate representation of the phase diagram of the model. Nonetheless, this
data shows that as opposed to the SU(3) and SU(4) cases discussed in the main
manuscript, for SU(2) the $R$ term alone does not destroy N\'eel order,
in fact it seems to favor it! This is seen because at ratios of
$Q/J_1$ that give a VBS, increasing $R$ drives the system back
into the N\'eel phase.

\section{Observables}

We follow previous work~\cite{block2013:fate} by defining an SU($N$) magnetic order parameter:
\begin{equation}
\label{eqn:Qab}
Q_{\alpha\beta}(\mathbf{r},\tau)=\left\{\begin{array}{ccc}\left(\Ket{\alpha}\Bra{\beta}\right)_{\mathbf{r},\tau}-\delta_{\alpha\beta}\frac{\hat{1}}{N}&,&\text{A sublattice}\\\left(\Ket{\beta}\Bra{\alpha}\right)_{\mathbf{r},\tau}-\delta_{\alpha\beta}\frac{\hat{1}}{N}&,&\text{B sublattice}\end{array}\right.,
\end{equation}
where $\alpha$ and $\beta$ vary over the $N$ colors.  We can then
define a magnetic order parameter as:
\begin{equation}
\label{eqn:chiN}
O^2_M\equiv\frac{1}{(N_s\beta)^2}\sum_{\mathbf{r},\mathbf{r}'}\int_0^\beta d\tau\int_0^\beta d\tau'\Braket{\text{T}_\tau Q_{\alpha\beta}(\mathbf{r},\tau)Q_{\beta\alpha}(\mathbf{r}',\tau')}.
\end{equation}

In a similar fashion we can define a VBS correlation function.  First we define the bond operator on a pair of nearest neighbor sites as follows:
\begin{equation}
\label{eqn:Bmu}
B^\mu(\mathbf{r},\tau)=\frac{1}{N}\mathcal{P}(\mathbf{r},\tau;\mathbf{r}+\hat{\mu},\tau),
\end{equation}
where
$\mathcal{P}_{ij}=\sum_{\alpha,\beta=1}^N\Ket{\alpha\alpha}_{ij}\Bra{\beta\beta}_{ij}$,
with spacetime locations of the two points given by the arguments.
The superscript $\mu$ denotes the bond type.  On the square lattice this index would run over $\mu=x,y$.  We can then study the correlations of these bond operators at different points in space and take the $\omega=0$ component:
\begin{widetext}
\begin{equation}
\label{eqn:Cmunu}
C^{\mu\nu}(\mathbf{r}-\mathbf{r}')\equiv\frac{1}{\beta^2}\int_0^\beta d\tau\int_0^\beta d\tau'\Braket{\text{T}_\tau B^\mu(\mathbf{r},\tau)B^\nu(\mathbf{r}',\tau')}-\Braket{B^\mu}\Braket{B^\nu}.
\end{equation}
\end{widetext}
The plaquette and columnar VBS patterns corresponds to a wavevector
$\mathbf{Q}=(\pi,0)$ and correlated bond type $\mu,\nu=x$.   By taking
the Fourier component of at this wavevector, we can check for a signal
in this VBS pattern.  This is how we define our VBS order parameter:
\begin{equation}
\label{eqn:chiV}
O^2_{\rm VBS}\equiv\frac{1}{N_s}\sum_{\mathbf{r}}C^{xx}(\mathbf{r})e^{i\mathbf{Q}\cdot\mathbf{r}}.
\end{equation}

Finally to create the VBS histograms we take a basis state and assign
1 to all bonds which have same SU($N$) colors on the sites that connect the
bonds and 0 to all bonds connecting different colors. We then Fourier transform all the $x$ directed
bonds to $(\pi,0)$ and call this $D_x$ and all the $y$ directed bonds
to $(0,\pi)$ and call this $D_y$. This gives us a value of $D_x$ and
$D_y$ for each basis state. We then histogram this data to get the
density plots shown in Fig.~\ref{fig:hist}.

\end{document}